%% file: main.tex
\documentclass[12pt]{article}
\usepackage{graphicx}
\usepackage{mathptmx}
\usepackage[intlimits,centertags]{amsmath}
\usepackage{amssymb,amsfonts}
\usepackage[pdftex]{hyperref}
\usepackage{aas_macros}
\usepackage{enumerate}
\usepackage[symbol]{footmisc}
\usepackage{xfrac}
\usepackage[table,xcdraw]{xcolor}
\usepackage{sectsty}
\usepackage[normalem]{ulem}
\usepackage{setspace}
\usepackage{wrapfig}
\usepackage{ragged2e}
\usepackage{combelow} 
\usepackage{etoolbox} 
\usepackage{cite}
\usepackage{fdsymbol}
\usepackage{pdfpages}
\usepackage[T1]{fontenc}
\usepackage{times}
\usepackage[compact]{titlesec}
\usepackage{geometry}
\usepackage[utf8]{inputenc}
\usepackage{enumitem,amssymb}
\usepackage{ragged2e}
\newlist{thematic}{itemize}{8}
\setlist[thematic]{label=$\square$}
\usepackage{pifont}

\definecolor{darkgreen}{rgb}{0,0.5,0}

\newif\ifastrophysical
\astrophysicaltrue
\newcommand{\onlyastrophysical}[1] 
{
  \ifastrophysical
  #1 
  \fi
}

\newif\iffundamental
\fundamentalfalse
\newcommand{\onlyfundamental}[1] 
{
  \iffundamental
  #1 
  \fi
}

\definecolor{header_color}{HTML}{225959}
\subsectionfont{\color{header_color}}

\begin{document}
\begin{titlepage}
\includepdf{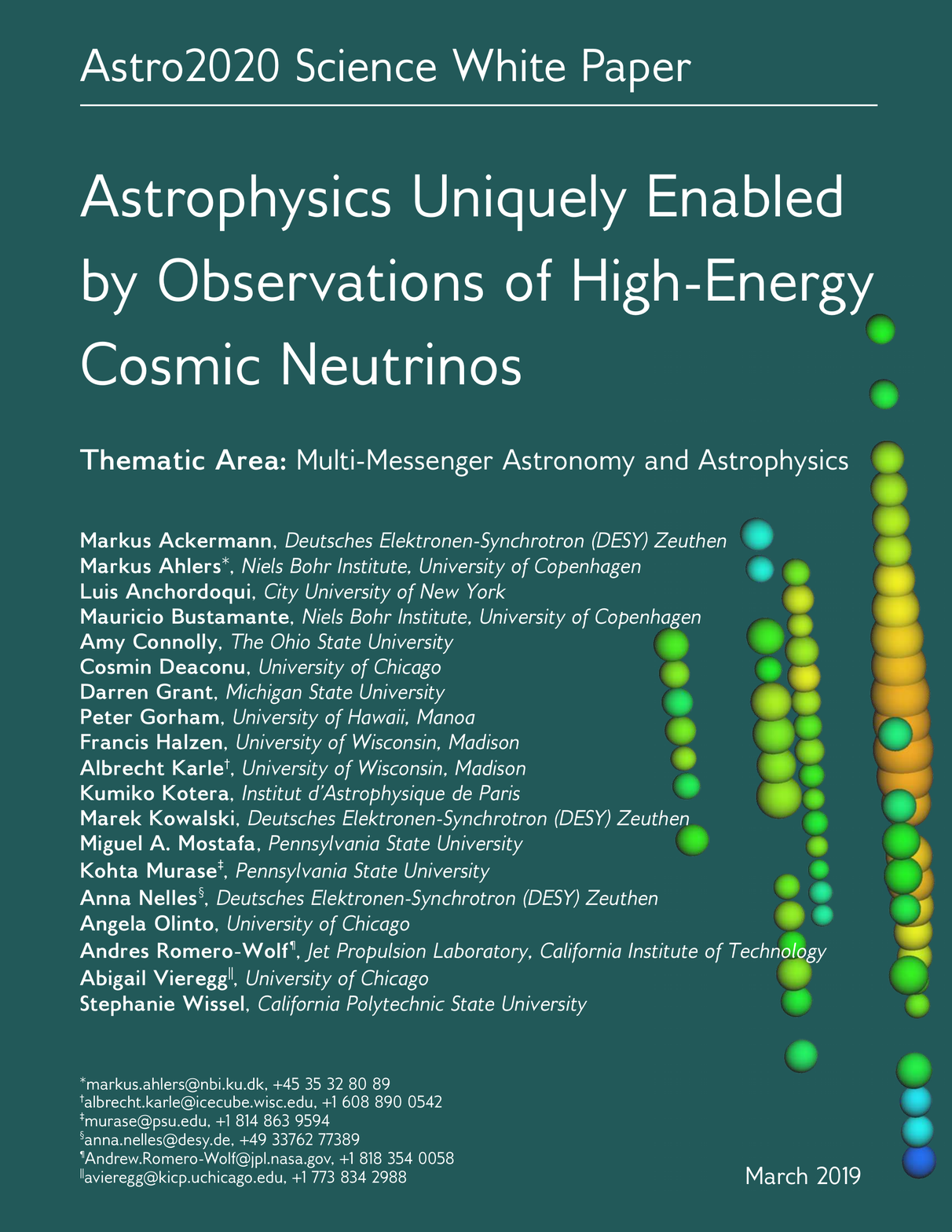}
\end{titlepage}

\pagenumbering{roman}

\begin{center}\textbf{Abstract}\end{center}
\justify
High-energy cosmic neutrinos carry unique information about the most energetic non-thermal sources in the Universe. This white paper describes the outstanding astrophysics questions that neutrino astronomy can address in the coming decade. A companion white paper discusses how the observation of cosmic neutrinos can address open questions in fundamental physics. Detailed measurements of the diffuse neutrino flux, measurements of neutrinos from point sources, and multi-messenger observations with neutrinos will enable the discovery and characterization of the most energetic sources in the Universe.\\

\begin{center}
\textbf{Endorsers}
\linebreak
\input{endorsers} 
\end{center}

\normalsize
\pagebreak
\subsection*{The Unique Tool of Neutrino Astronomy}

\pagenumbering{arabic}

Neutrino astronomy allows us to discover and characterize the most energetic non-thermal sources in the Universe. Despite observations of cosmic rays (charged nuclei), which reach energies that are ten million times higher than those achievable in the Large Hadron Collider~\cite{Beatty:2009zz,Evans:2008zzb}, and observations of $\gamma$-rays~\cite{Funk:2015ena} and astrophysical neutrinos~\cite{Aartsen:2013bka,Aartsen:2013jdh,Aartsen:2014gkd,Kopper:2015vzf,Aartsen:2016xlq,Haack:2017dxi,Kopper:2017zzm}, we do not yet know where or how these particles are accelerated. Neutrino astronomy is a key to directly answering the question of how particles are accelerated to these extreme energies.
Cosmic rays can collide with gas and radiation 
in their sources or while propagating 
over cosmic distances 
until they reach Earth.  A ``smoking-gun'' signal of such interactions is the production of high-energy neutrinos. 

Astrophysical neutrinos provide insight into source characteristics not accessible through the observation of other messengers. Due to their low cross sections, neutrinos can escape dense astrophysical environments that are opaque to photons. In contrast to $\gamma$-rays, neutrinos travel almost unimpeded through the Universe, allowing direct observation of their sources at high redshifts with sub-degree-scale pointing. Unlike cosmic rays, neutrinos are not deflected in magnetic fields and can be observed in spatial and temporal coincidence with photons and gravitational waves~\cite{Abbott:2016blz,TheLIGOScientific:2017qsa}, which is a key prerequisite to reap the scientific rewards of multi-messenger astronomy. 
In addition, neutrinos come in different flavors --- electron, muon, and tau neutrinos ($\nu_e$, $\nu_\mu$, \& $\nu_\tau$) --- and the flavor ratios observed at Earth give insight into the environment of cosmic-ray sources.

\begin{figure}\centering
\includegraphics[width=0.7\linewidth,viewport=5 30 645 350,clip=false]{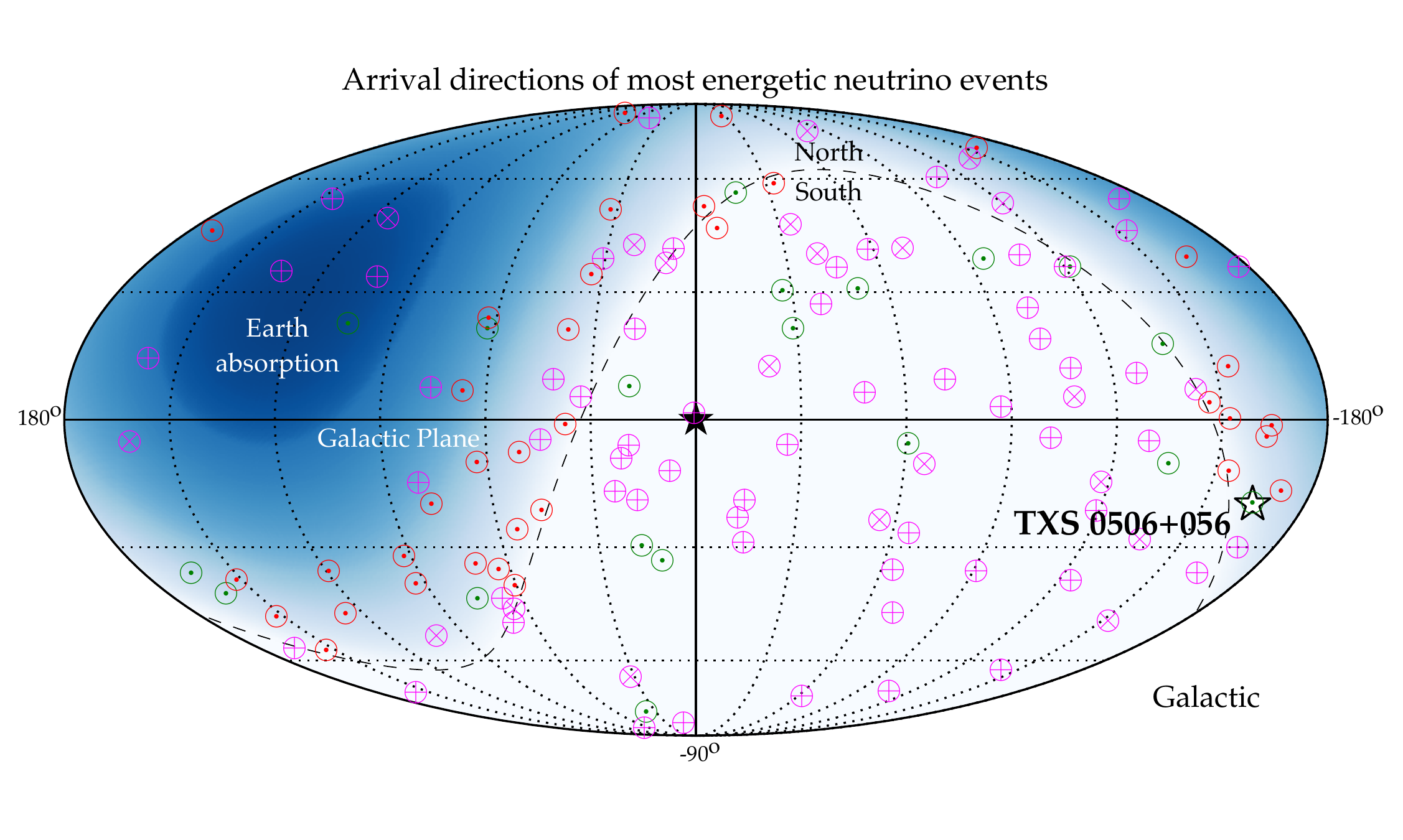}\\[-0.3cm]
\caption[]{Arrival directions of neutrino events from IceCube. Shown are upgoing track events~\cite{Aartsen:2016xlq,Haack:2017dxi} (\textcolor{red}{$\odot$}), the high-energy starting events (HESE) (tracks \textcolor{magenta}{$\otimes$} and cascades \textcolor{magenta}{$\oplus$})~\cite{Aartsen:2014gkd,Kopper:2015vzf,Kopper:2017zzm}, and additional track events published as public alerts (\textcolor{darkgreen}{$\odot$})~\cite{Smith:2012eu,GCN}. The blue-shaded region indicates where the Earth absorption of 100-TeV neutrinos becomes important. The dashed line indicates the equatorial plane. We also indicate the location of the blazar TXS 0506+056 ($\medwhitestar$).}
\label{fig:skymap}
\end{figure}

The last decade ushered in high-energy neutrino astronomy, with the discovery of an astrophysical neutrino flux in the 10~TeV~--~10~PeV energy range~\cite{Aartsen:2013bka,Aartsen:2013jdh,Aartsen:2014gkd,Kopper:2015vzf,Aartsen:2016xlq,Haack:2017dxi,Kopper:2017zzm}. 
The arrival directions of the most energetic neutrinos are shown in Fig.~\ref{fig:skymap} 
and are consistent with a uniform distribution across the sky after accounting for detector acceptance. Neutrino emission at the observed flux level has been predicted from a variety of source classes, including $\gamma$-ray bursts, blazars, starburst galaxies, galaxy clusters, and others (see, {\it e.g.} \cite{Halzen:2016gng,Ahlers:2018fkn}). Recently, coincident observations of neutrinos and $\gamma$-rays from the blazar TXS~0506+056 presented evidence of the first extragalactic neutrino source~\cite{IceCube:2018cha,IceCube:2018dnn}. However, this cannot be the entire story: multiple independent analyses indicate that only a fraction of the diffuse neutrino flux can come from $\gamma$-ray blazars~\cite{Murase:2016gly,Ando:2017xcb,Aartsen:2016lir,Neronov:2016ksj,Huber:2017wxt,Hooper:2018wyk}. 

In the next decade, 
the development, construction, and operation of multiple
neutrino detectors that cover complementary parts of the sky, have a wide range of neutrino energies, and have sensitivity to different flavors, will disentangle the complexities of the neutrino sky.   
Real-time multi-messenger campaigns, in collaboration with multi-wavelength (radio to $\gamma$-ray) and gravitational-wave astronomers, could prove crucial in unveiling the sources of the most energetic particles and the acceleration mechanisms at work. Neutrinos would provide insights into the physics of stellar explosions, compact object mergers, and relativistic jets, as well as particle acceleration processes.

\subsection*{Discovering and Characterizing the Most Energetic Sources in the Universe}

The goal of discovering the most energetic non-thermal sources in the Universe can be approached through multiple observational avenues. Detailed observations of all cosmic messengers, including neutrinos, are needed to fully understand the processes at work. Precision measurements of the diffuse neutrino spectrum will shed light on the physics of the most energetic non-thermal sources and their host environments. High-resolution neutrino data from observatories with deep exposure and wide sky coverage will allow us to identify the source population(s) responsible for the diffuse neutrino emission. Combining observations of these neutrinos and other cosmic messengers will provide the optimal strategy of identifying these sources and determining the governing physics.

\begin{figure}[t]\centering
\includegraphics[width=0.41\linewidth]{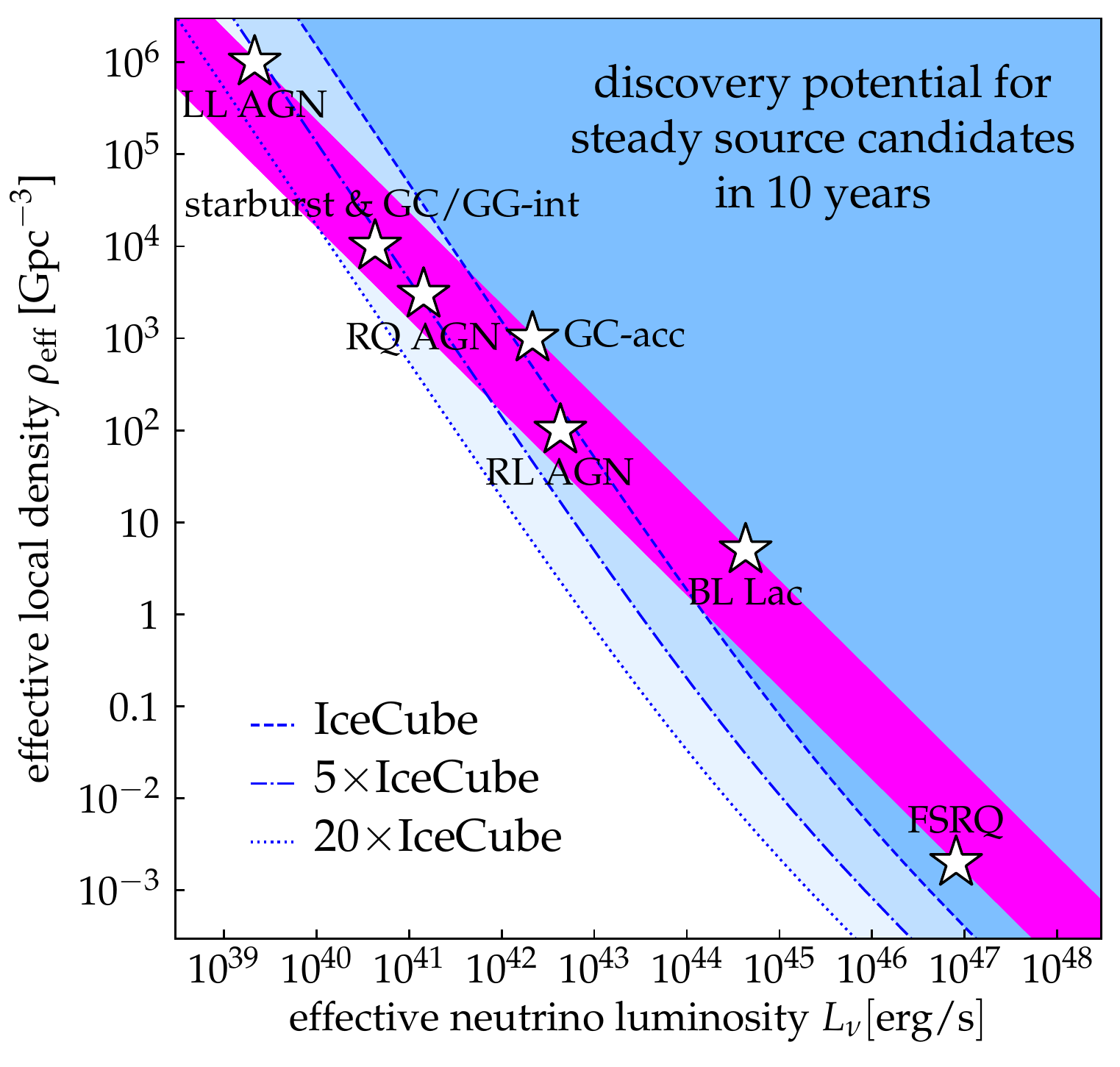}\hspace{0.5cm}\includegraphics[width=0.41\linewidth]{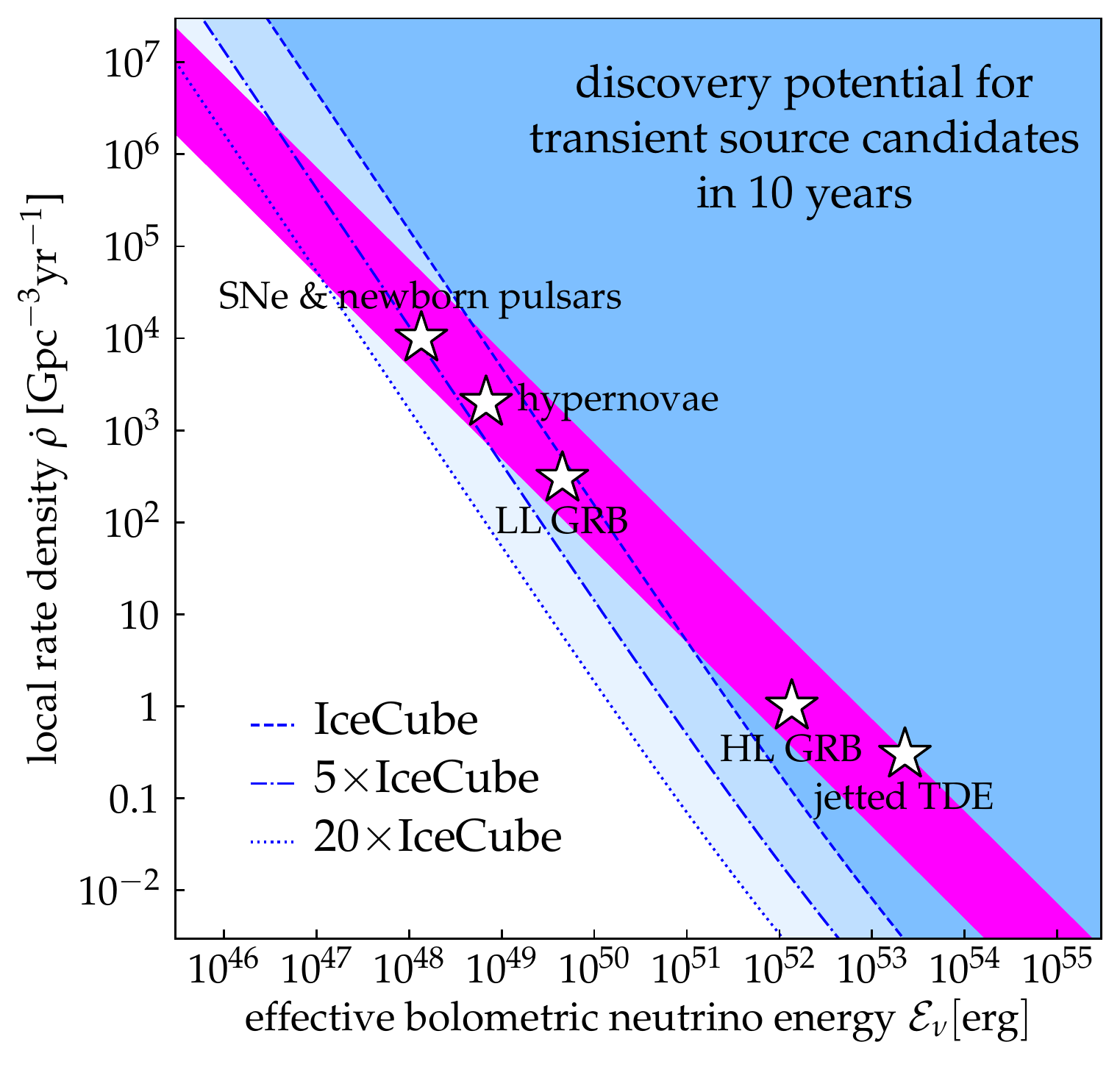}\\[-0.3cm]
\caption[]{
{\bf Left:} Comparison of the diffuse neutrino emission (solid magenta band) to the effective local density and luminosity of extragalactic neutrino source populations. We indicate several candidate populations ($\medwhitestar$) by the required neutrino luminosity to account for the full diffuse flux~\cite{Murase:2016gly} (see also \cite{Silvestri:2009xb}). The lower (upper) edge of the band assumes rapid (no) redshift evolution. The dark-blue-shaded region indicates IceCube's discovery potential of the closest source of the population ($E^2\phi_{\nu_\mu+\bar\nu_\mu}\simeq 10^{-12}~{\rm TeV}/{\rm cm}^2/{\rm s}$ in the Northern Hemisphere~\cite{Aartsen:2018ywr}). {\bf Right:} The same comparison for transient neutrino sources parametrized by their local density rate and bolometric energy~\cite{Murase:2018utn}. The discovery potential of the closest source is based on 10 years of livetime ($E^2F_{\nu_\mu+\bar\nu_\mu}\simeq 0.1~{\rm GeV}/{\rm cm}^2$ in the Northern Hemisphere~\cite{Meagher:2017htr}).
}\label{fig:diffuse_vs_PS}
\end{figure}

The current lack of established neutrino point sources --- despite a firm detection of a diffuse neutrino flux --- indicates a population of weak extragalactic sources. This is illustrated in Fig.~\ref{fig:diffuse_vs_PS}, which shows a parametrization of the diffuse flux (magenta bands) in terms of the local density and luminosity of steady source populations~\cite{Murase:2016gly} (left plot) or local density rate and bolometric energy for transient source populations~\cite{Murase:2018utn} (right plot). 
The lack of neutrino sources after ten years of observations by IceCube translates into the dark-blue shaded exclusion regions. Source populations with sufficiently large local densities --- like starburst galaxies~\cite{Loeb:2006tw,delPozo:2009mh,Murase:2013rfa,Liu:2013wia,Katz:2013ooa,Tamborra:2014xia,Anchordoqui:2014yva,Chang:2014hua,Chakraborty:2015sta,Senno:2015tra}, galaxy clusters and groups~\cite{Murase:2008yt,Kotera:2009ms,Murase:2013rfa,Fang:2017zjf}, low-luminosity AGN~\cite{Kimura:2014jba}, radio-quiet AGN~\cite{AlvarezMuniz:2004uz,Stecker:2013fxa,Kalashev:2015cma}, or star-forming galaxies with AGN outflows~\cite{Tamborra:2014xia,Wang:2016vbf,Lamastra:2017iyo,Liu:2017bjr,Padovani:2018hfm} --- or with high local rate densities --- like (extragalactic) jet-powered SNe including hypernovae~\cite{Murase:2013ffa,Tamborra:2015fzv,Senno:2015tsn,He:2018lwb} and interaction-powered SNe~\cite{Zirakashvili:2015mua,Petropoulou:2017ymv} --- are presently consistent with the observations. 
Observatories with improvements in point-source sensitivity over current detectors would greatly expand the discovery potential for the brightest sources of these candidate populations (see Fig.~\ref{fig:diffuse_vs_PS}) and other candidate sources like TXS 0506+056.

\begin{figure}[t]\centering
\includegraphics[width=0.78\linewidth]{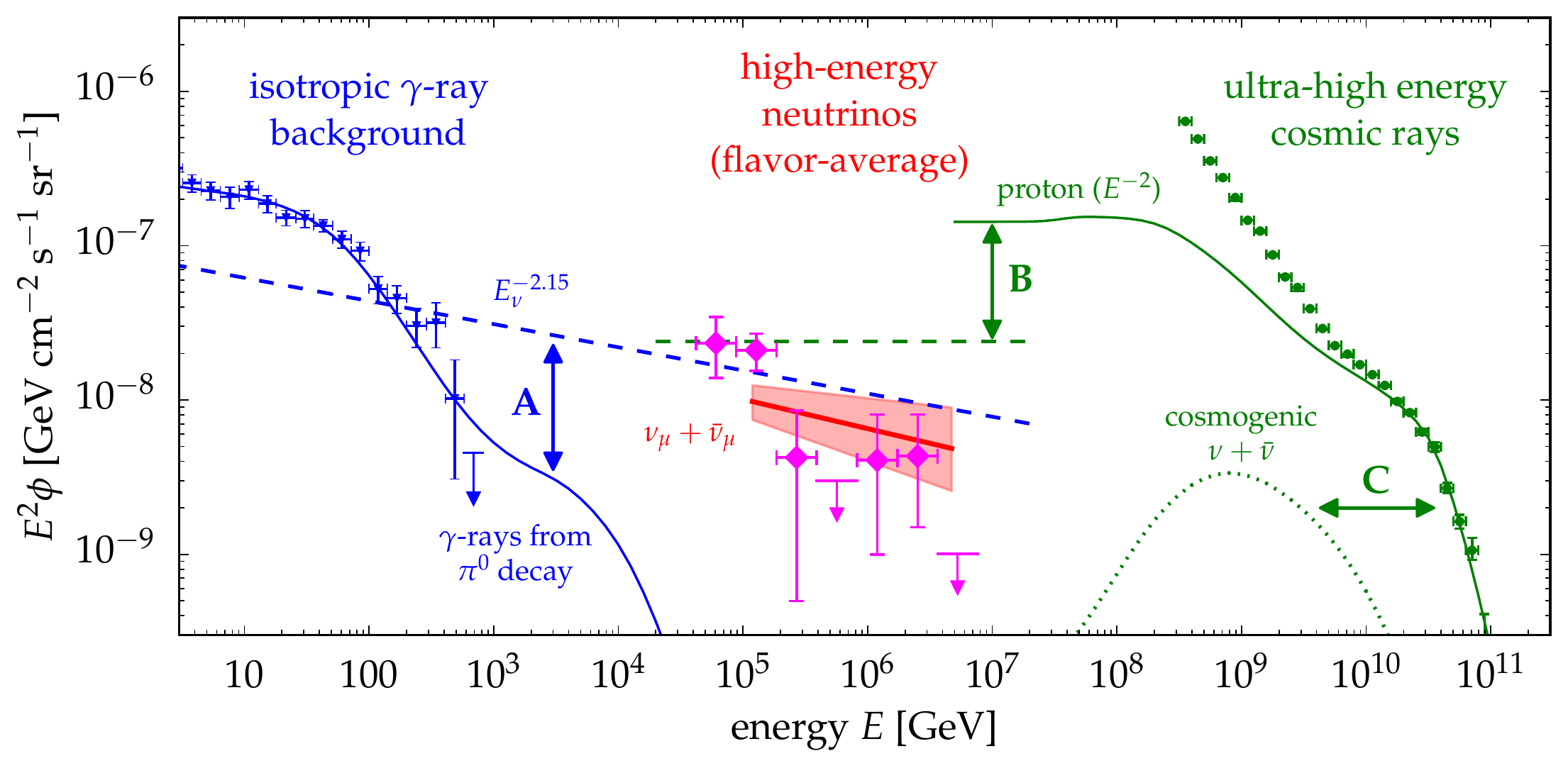}\\[-0.4cm]
\caption[]{The flux ($\phi$) of neutrinos, per flavor \cite{Haack:2017dxi,Kopper:2017zzm} (red and magenta data) compared to the flux of unresolved extragalactic $\gamma$-ray emission~\cite{Ackermann:2014usa} (blue data) and UHE cosmic rays~\cite{Valino:2015zdi} (green data). We highlight two neutrino limits (dashed lines) based on multi-messenger models~\cite{Murase:2013rfa,Murase:2016gly}.}\label{fig:diffuse}
\end{figure}

Current measurements of the isotropic neutrino flux ($\phi$) are shown in Fig.~\ref{fig:diffuse}, along with the observed isotropic $\gamma$-ray background (IGB) and the UHE cosmic-ray flux.  The correspondence among the energy densities, proportional to $E^2\phi$, observed in neutrinos, $\gamma$-rays, and cosmic rays suggests a strong multi-messenger relationship that offer intriguing prospects for deeper observations with a new generation of instruments.

{\bf A)} The simultaneous production of neutral and charged pions in cosmic-ray interactions suggests that the sources of high-energy neutrinos could also be strong 10~TeV~--10~PeV $\gamma$-ray emitters. For extragalactic scenarios, this $\gamma$-ray emission is not directly observable because of the strong absorption of photons by $e^+e^-$ pair production in extragalactic background photons. High-energy $\gamma$-rays initiate electromagnetic cascades of repeated inverse-Compton scattering and pair production that eventually contribute to the diffuse $\gamma$-rays below 100~GeV, which provides a theoretical upper limit to the diffuse neutrino flux~\cite{Berezinsky:1975zz,Mannheim:1998wp}.
The detected flux of $>100$~TeV neutrinos with the hadronuclear origin is saturated by the diffuse $\gamma$-ray data~\cite{Murase:2013rfa} (see blue lines in Fig.~\ref{fig:diffuse}). 
Intriguingly, several IceCube analyses~\cite{Aartsen:2014muf,Kopper:2017zzm} show an excess of neutrinos below 100~TeV, indicating that the sources are opaque to $\gamma$-rays, as expected, {\it e.g.}, for intense X-ray and soft $\gamma$-ray sources~\cite{Murase:2015xka}.

{\bf B)} Precision measurements of the neutrino flux can test the idea of cosmic particle unification, in which sub-TeV $\gamma$-rays, PeV neutrinos, and UHE cosmic rays can be explained simultaneously~\cite{Murase:2016gly,Fang:2017zjf,Kachelriess:2017tvs,Resconi:2017zkg}.
If the neutrino flux is related to the sources of UHE cosmic rays, then there is a different theoretical upper limit (the dashed green line in Fig.~\ref{fig:diffuse}) to the neutrino flux~\cite{Waxman:1998yy,Bahcall:1999yr}. 
UHE cosmic ray sources can be embedded in environments that act as ``cosmic-ray reservoirs'' where magnetic fields trap cosmic rays with energies far below the highest cosmic-ray energies.
The trapped cosmic rays collide with gas and produce a flux of $\gamma$-rays and neutrinos. 
The measured IceCube flux is consistent with predictions of some of these models~\cite{Loeb:2006tw,Murase:2008yt,Kotera:2009ms}; see, however, \cite{Anchordoqui:2018qom}. 

{\bf C)} The attenuation of UHE cosmic rays through resonant interactions with cosmic microwave background photons results in the production of UHE neutrinos. This mechanism, first pointed out by Greisen, Zatsepin and Kuzmin~\cite{Greisen:1966jv,Zatsepin:1966jv} (GZK), causes a suppression of the UHE cosmic ray proton flux beyond $5\times10^{10}$~GeV~\cite{Greisen:1966jv,Zatsepin:1966jv} and gives rise to a flux of UHE neutrinos~\cite{Beresinsky:1969qj}, not yet detected, shown in Fig.~\ref{fig:diffuse}.
The observation of these cosmogenic neutrinos at $\sim$EeV, or a stringent upper limit on their flux, will severely restrict models of acceleration, source evolution, cosmic ray composition, and transition from Galactic to extragalactic components, and serve as a complement to cosmic-ray measurements to limit possible sources ({\it e.g.}, \cite{Beresinsky:1969qj,Berezinsky:1975zz,Stecker:1978ah,Hill:1983xs,Yoshida:1993pt,Engel:2001hd,Anchordoqui:2007fi,Takami:2007pp,Ahlers:2009rf,Ahlers:2010fw,Kotera:2010yn,Yoshida:2012gf,Ahlers:2012rz,Aloisio:2015ega,Heinze:2015hhp,Romero-Wolf:2017xqe,AlvesBatista:2018zui,Moller:2018isk,vanVliet:2019nse,Heinze:2019jou}).  

The strong correspondence of high-energy messengers --- suggested by the diffuse data in Fig.~\ref{fig:diffuse} --- provides excellent motivation for multi-messenger observations.
Linking together observations of multiple messengers in time and space 
will allow direct correlation of neutrino sources with specific sources of $\gamma$-rays and offers a wealth of information that is not available with neutrino astronomy alone. The most successful example so far is the multi-messenger flare of TXS 0506+056~\cite{IceCube:2018dnn}, which demonstrated the feasibility of neutrino-triggered follow-up campaigns. However, there is no simple concordance picture of neutrino emission from this source~\cite{Ahnen:2018mvi,Keivani:2018rnh,Murase:2018iyl,Sahakyan:2018voh,Padovani:2018acg,Cerruti:2018tmc,Kun:2018zin,Gao:2018mnu,Righi:2018hhu,Liu:2018utd,Zhang:2018xrr,Wang:2018zln,Reimer:2018vvw,Rodrigues:2018tku,Padovani:2019xcv}, so further studies are required to establish blazar flares as sources of high-energy neutrinos. 

High-energy neutrino observations will allow us to investigate the rich diversity of stellar explosions --- ranging from core-collapse SNe~\cite{Murase:2010cu,Murase:2017pfe}, over trans-relativistic SNe~\cite{Katz:2011zx,Kashiyama:2012zn} associated with low-luminosity $\gamma$-ray bursts~\cite{Murase:2006mm,Gupta:2006jm,Murase:2008mr,Senno:2015tsn,Zhang:2017moz,Boncioli:2018lrv,Zhang:2018agl}, jet-powered SNe~\cite{Meszaros:2001ms,Razzaque:2003uv,Razzaque:2004yv,Ando:2005xi,Iocco:2007td,Murase:2013ffa}, and wind-powered SNe~\cite{Murase:2009pg,Fang:2013vla,Fang:2018hjp}, 
to $\gamma$-ray bursts with ultra-relativistic jets~\cite{Waxman:1997ti,Rachen:1998fd,Dermer:2003zv,Guetta:2003wi,Razzaque:2003uw,Murase:2005hy,Murase:2008sp,Wang:2008zm,Baerwald:2010fk,Ahlers:2011jj,Murase:2011cx,Li:2011ah,Hummer:2011ms,He:2012tq,Zhang:2012qy,Liu:2012pf,Gao:2013fra,Petropoulou:2014awa,Petropoulou:2014lja,Bustamante:2014oka,Wang:2007xj,Murase:2008mr,Calvez:2010uh,Globus:2014fka,Biehl:2017zlw,Paczynski:1994uv,Bartos:2013hf,Murase:2013hh,Murase:2006dr,Waxman:1999ai,Dermer:2000yd,Murase:2007yt,Razzaque:2013dsa}. 
Neutrino-triggered follow-up searches~\cite{Murase:2006mm,Kowalski:2007xb,Aartsen:2015trq,Adrian-Martinez:2015nin,Aartsen:2017snx,Kankare:2019bzi} and stacking analyses~\cite{Abbasi:2012zw,Senno:2017vtd,Esmaili:2018wnv} are in reach of testing the predictions. 
Other candidate transient neutrino sources are jetted tidal disruption events (TDE)~\cite{Wang:2015mmh,Dai:2016gtz,Senno:2016bso,Lunardini:2016xwi,Zhang:2017hom,Biehl:2017hnb,Guepin:2017abw}, flaring flat spectrum quasars~\cite{Atoyan:2001ey,Dermer:2014vaa,Kadler:2016ygj,Halzen:2016uaj,Kun:2016bnk}, and compact object mergers~\cite{Kimura:2017kan,Fang:2017tla,Biehl:2017qen,Kimura:2018vvz,Xiao:2016man,Murase:2016etc,Kotera:2016dmp,Moharana:2016xkz}. The latter are also intriguing targets for coincident detection of neutrinos and gravitational waves~\cite{Kimura:2017kan,Fang:2017tla}, and models have been constrained for the recent merger event GW170817~\cite{ANTARES:2017bia}.
Steady emission from Galactic neutrino sources could contribute a fraction of the observed diffuse flux~\cite{Ahlers:2013xia,Anchordoqui:2014rca,Adrian-Martinez:2014wzf,Adrian-Martinez:2015ver,Adrian-Martinez:2016fei,Aartsen:2017ujz,Aartsen:2018ywr,Albert:2018vxw} ({\it e.g.}, stellar explosion remnants~\cite{Kistler:2006hp,Kappes:2006fg,Torres:2010uk,Vissani:2011vg,Fox:2013oza,Ahlers:2013xia,Mandelartz:2013ht,Andersen:2017yyg}, $\gamma$-ray binaries~\cite{Levinson:2001as,Distefano:2002qw,Torres:2006ub,Anchordoqui:2014rca}, star-forming regions~\cite{Anchordoqui:2006pe,Anchordoqui:2009tya,Ahlers:2013xia,Gonzalez-Garcia:2013iha,Bykov:2015nta}, Galactic center and ridge regions~\cite{Neronov:2013lza,Razzaque:2013uoa,Ahlers:2013xia,Lunardini:2013gva,Supanitsky:2013ooa,Bai:2014kba,Fujita:2015xva,Anchordoqui:2016dcp,Fang:2017vlg}, diffuse emission~\cite{Stecker:1978ah,Anchordoqui:2013qsi,Ahlers:2013xia,Joshi:2013aua,Neronov:2014uma,Kachelriess:2014oma,Gaggero:2015xza,Denton:2017csz}, and quasi-isotropic halo emission~\cite{Ahlers:2013xia,Taylor:2014hya,Kalashev:2016euk,Liu:2018gyj,Blasi:2019obb}).

\subsection*{Observatory Requirements to Achieve the Science Goals}
 
\begin{figure}\centering
\begin{minipage}[c][6.8cm][t]{0.495\linewidth}\centering
\rule{0pt}{0pt}\\\includegraphics[height=6.8cm]{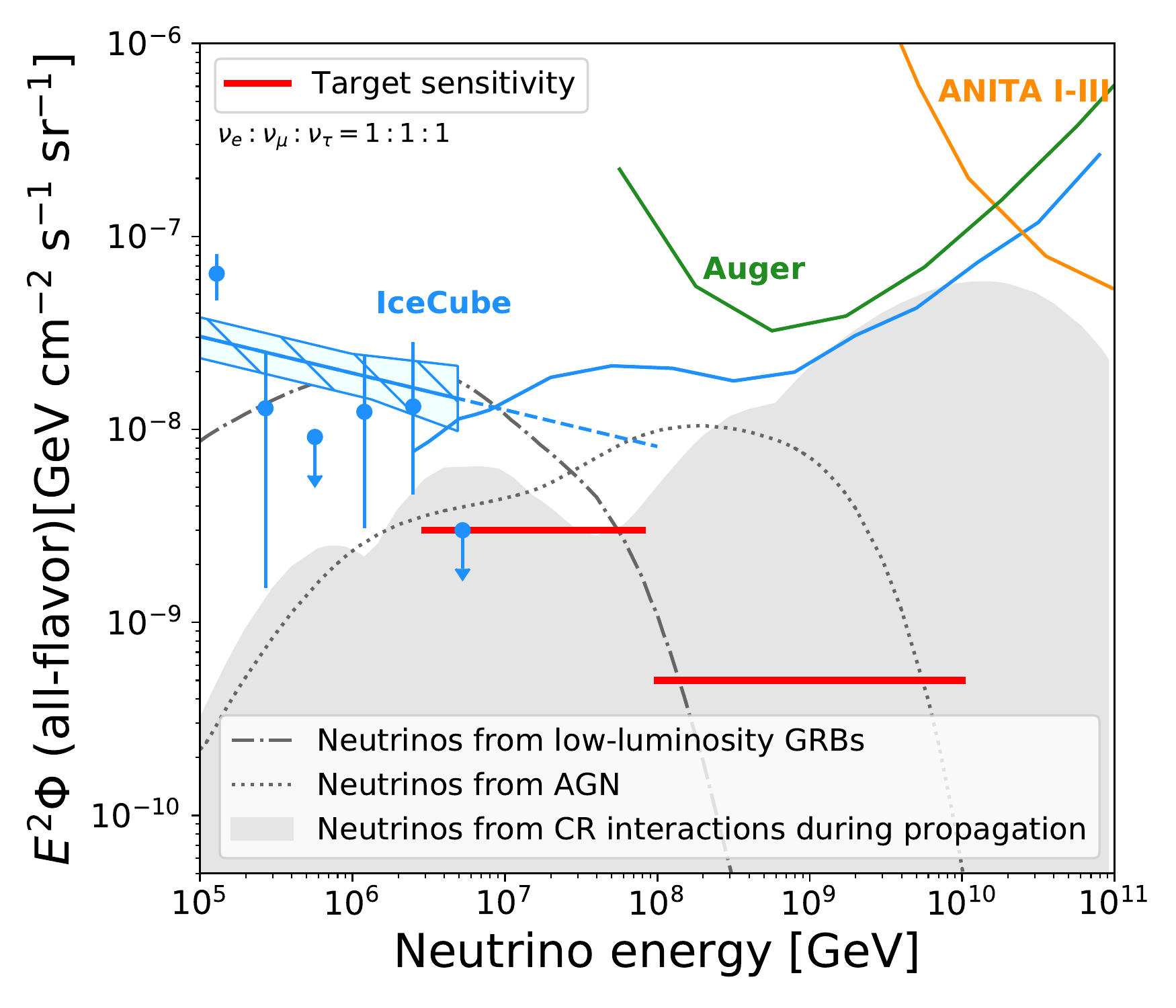}
\end{minipage}
\begin{minipage}[c][6.8cm][t]{0.459\linewidth}\centering
\rule{0pt}{0pt}\\[0.26cm]\includegraphics[height=5.66cm]{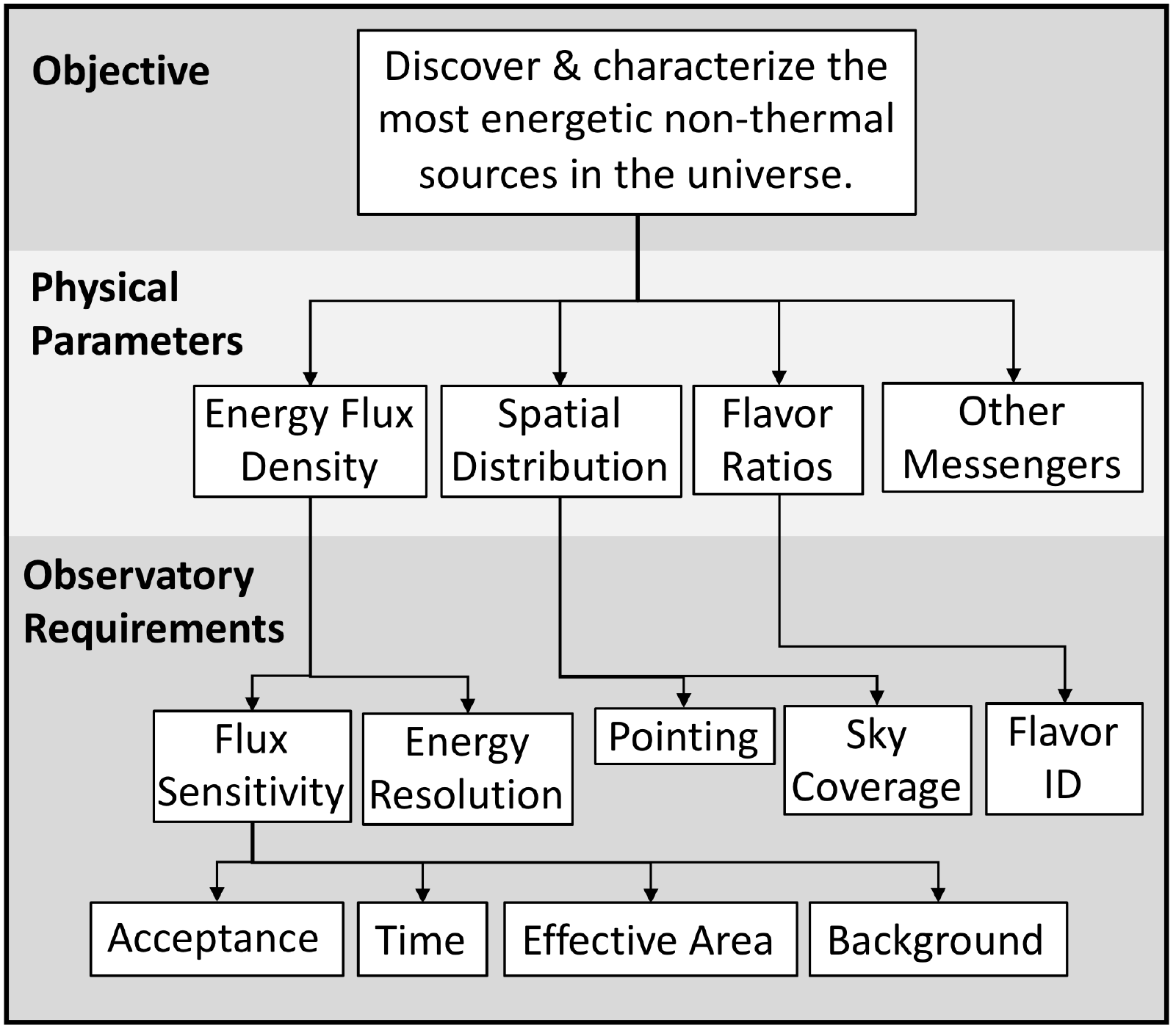}
\end{minipage}\\[-0.2cm]
\caption[]{{\bf Left:} Current experimental limits and detections in neutrino astronomy from IceCube \cite{Haack:2017dxi,Kopper:2017zzm, Aartsen:2018vtx}, the Pierre Auger Observatory \cite{Aab:2015kma}, and ANITA \cite{Allison:2018cxu}. Also shown are low-luminosity GRB~\cite{Boncioli:2018lrv} (see \cite{Murase:2013ffa,Senno:2015tsn} for similar spectra) 
and AGN models~\cite{Murase:2014foa}, and an extrapolation of the IceCube flux, which suggests target sensitivities for the next observatories. {\bf Right:} Observatory requirements for neutrino astronomy targeting different physical parameters.}\label{fig4}
\end{figure}

Meeting these science goals requires measurements of the neutrino flux density, the neutrino spatial distribution, neutrino flavor ratios, and requires linking neutrino observations with observations of complementary astrophysical messengers (see Fig.~\ref{fig4}).  This flows down to measurement requirements for astrophysical neutrino observatories in the coming decade. 
 Measuring neutrino point-source and diffuse energy flux densities will require large detector arrays. 
As detector effective areas increase, it is imperative to maintain low backgrounds to achieve improved sensitivity. 

The spatial distribution and clustering of high-energy neutrinos across the sky are key observables for revealing their origins.  Source catalogue correlations require sub-degree pointing resolution~\cite{Ahlers:2014ioa,Murase:2016gly,Fang:2016hop,Bartos:2016wud}. At energies between 10~TeV~--~10~PeV, an order-of-magnitude improvement in point-source sensitivity will be needed to discover neutrino point sources consistent with the flux discovered by IceCube~\cite{Ahlers:2014ioa,Murase:2016gly} (see Fig.~\ref{fig:diffuse_vs_PS}).

For energies 1~--~100~PeV, an order-of-magnitude improvement in the diffuse neutrino flux sensitivity is needed to identify a break, a cutoff, or a new component, to probe diffuse model predictions ({\it e.g.},~\cite{Murase:2006mm,Murase:2013ffa,Murase:2013rfa,Murase:2014foa,Kistler:2016ask,Anchordoqui:2018qom}).
An energy resolution of $\Delta E/E \lesssim 0.5$ is required to distinguish between model fluxes.
At $E\gtrsim$100 PeV, two orders of magnitude improvement in the diffuse flux sensitivity is needed to test whether particle accelerators in the Universe have similar characteristics to those in our local ($\lesssim$100 Mpc) neighborhood~\cite{Romero-Wolf:2017xqe,AlvesBatista:2018zui, Wittkowski:2018giy} (see left panel in Fig.~\ref{fig4}). %

The observed ratios of neutrino flavors and the observed ratio of neutrinos to anti-neutrinos 
provides a complementary probe of neutrino production mechanisms and the physical conditions at the sources~\cite{Anchordoqui:2004eb, Ackermann:2019cxh}. 
The identification of per-flavor neutrino fluxes require large event statistics at observatories that can identify flavor-specific signals.
 
Multi-messenger observations that include neutrinos will provide additional information to localize and characterize sources. This requires the ability to both send and receive real-time alerts to and from other telescopes and gravitational wave detectors~\cite{Smith:2012eu}.

We advocate a staged, multi-observatory approach that will extend the science reach of neutrino observatories throughout the next decade.  Complementary approaches that target a broad range of energies (from the TeV scale up through the EeV scale) are useful for determining the nature of these high energy sources.
Multiple detectors will be required to measure the per-flavor flux, cover large fractions of the sky, and provide complementary sky coverage in the North and South to increase the exposure for multi-messenger observations.  Multiple detectors, preferably with different detection mechanisms to reduce systematic effects, will also be required to confirm results. We advocate for strong cooperation with multi-messenger partners across the electromagnetic spectrum, cosmic rays, and gravitational waves.  
 
\clearpage

\bibliographystyle{utphysmod.bst}
\bibliography{references}

\end{document}

%% file: endorsers.tex
 \newcounter{AffiliationCount} 
 \def\affiliations{} 

 \newcommand{\newAff}[2]
 {
     \stepcounter{AffiliationCount}
     \xdef#1{$^{\theAffiliationCount}$}
     \appto\affiliations {{#1\textit{\mbox{#2}\quad}} }
 }

\def\affiliations{\footnotesize}

\newAff{\atUCI}{University of California, Irvine} 
\newAff{\atIOPB}{Institute of Physics, Bhubaneswar} 
\newAff{\atULB}{Universit\'e Libre de Bruxelles} 
\newAff{\atClemson}{Clemson University} 
\newAff{\atINFN}{Istituto Nazional di Fisica Nucleare (INFN)}
\newAff{\atGSSI}{Gran Sasso Science Institute (GSSI)}
\newAff{\atCompostela}{Universidade de Santiago de Compostela} 
\newAff{\atSP}{Universidade de S\~ao Paulo} 
\newAff{\atChungbuk}{Chungbuk National University}
\newAff{\atMarquette}{Marquette University}
\newAff{\atAmsterdam}{Universiteit van Amsterdam} 
\newAff{\atErlangen}{Friedrich-Alexander-Universit\"at Erlangen-N\"urnberg} 
\newAff{\atSorbonne}{Sorbonne Universit\'e} 
\newAff{\atBerne}{Universit\'e de Berne}  
\newAff{\atTokyo}{University of Tokyo} 
\newAff{\atNBI}{Niels Bohr Institute, University of Copenhagen}
\newAff{\atRWTHAachen}{Rheinisch-Westf\"alische Technische Hochschule Aachen}
\newAff{\atPSU}{Pennsylvania State University}
\newAff{\atSDSM}{South Dakota School of Mines and Technology}
\newAff{\atValencia}{Institut de F\'isica Corpuscular, Universitat de Val\`encia} 
\newAff{\atUW}{University of Wisconsin, Madison}
\newAff{\atFlorida}{University of Florida}
\newAff{\atOSU}{The Ohio State University}
\newAff{\atMelbourne}{University of Melbourne}
\newAff{\atAdelaide}{University of Adelaide}
\newAff{\atRochester}{University of Rochester}
\newAff{\atUtah}{University of Utah}
\newAff{\atPadova}{Universit\`a degli Studi di Padova}
\newAff{\atDESYZeuthen}{Deutsches Elektronen-Synchrotron (DESY) Zeuthen}
\newAff{\atTurin}{Universit\`a degli Studi di Torino}
\newAff{\atNASAMarshall}{NASA Marshall Space Flight Center}
\newAff{\atINFNTurin}{Istituto Nazional di Fisica Nucleare (INFN), Sezione di Torino}
\newAff{\atIPNO}{Institut de Physique Nucl\'eaire d'Orsay (IPNO), Universit\'e Paris-Sud, Universit\'e Paris-Saclay}
\newAff{\atUMD}{University of Maryland, College Park} 
\newAff{\atINNPPP}{Institut National de Physique Nucl\'eaire et de Physique des Particules (IN2P3)}
\newAff{\atZagreb}{University of Zagreb}
\newAff{\atUppsala}{Uppsala Universitet}
\newAff{\atCFA}{Center for Astrophysics, Harvard \& Smithsonian}
\newAff{\atINFNBari}{Istituto Nazional di Fisica Nucleare (INFN), Sezione di Bari}
\newAff{\atGoddard}{NASA Goddard Space Flight Center}
\newAff{\atCatania}{Universit\`a degli Studi di Catania}
\newAff{\atFSU}{Florida State University}
\newAff{\atMPI}{Max-Planck-Institut f\"ur Kernphysik, Heidelberg} 
\newAff{\atWitwatersrand}{University of the Witwatersrand}
\newAff{\atNTU}{National Taiwan University}
\newAff{\atAnnaba}{Badji Mokhtar University of Annaba}
\newAff{\atCUNY}{City University of New York}
\newAff{\atVUB}{Vrije Universiteit Brussels}
\newAff{\atMarseille}{Centre de Physique des Particules de Marseille (CPPM)}
\newAff{\atUCL}{University College London}
\newAff{\atHongKong}{The University of Hong Kong}
\newAff{\atTata}{Tata Institute of Fundamental Research, Mumbai (TIFR)}
\newAff{\atNW}{Northwestern University}
\newAff{\atRio}{Universidade Federal do Rio de Janeiro}
\newAff{\atNijmegen}{Radboud Universiteit Nijmegen}
\newAff{\atNikhef}{Nikhef}
\newAff{\atIAP}{Institut d'Astrophysique de Paris}
\newAff{\atBNL}{Brookhaven National Laboratory} 
\newAff{\atMSU}{Michigan State University}
\newAff{\atChicago}{University of Chicago}
\newAff{\atMunster}{Westf\"alische Wilhelms-Universit\"at M\"unster}
\newAff{\atLund}{Lunds Universitet}
\newAff{\atLMU}{Ludwig-Maximilians-Universit\"at M\"unchen}
\newAff{\atRIKEN}{RIKEN}
\onlyfundamental{\newAff{\atKings}{King's College London}}
\newAff{\atKIT}{Karlsruher Institut f\"ur Technologie}
\newAff{\atColoradoMines}{Colorado School of Mines} 
\newAff{\atPontificiaRio}{Pontificia Universidade Catolic\'a do Rio de Janeiro}
\newAff{\atStanford}{Stanford University}
\newAff{\atOKState}{Oklahoma State University} 
\newAff{\atStockholm}{Stockholm Universitet}
\newAff{\atKyoto}{Kyoto University}
\newAff{\atUAM}{Instituto de F\'isica Te\'orica UAM-CSIC}
\newAff{\atPeru}{Pontificia Universidad Cat\'olica del Per\'u}
\newAff{\atDelaware}{Bartol Research Institute, University of Delaware}

\newAff{\atLaPlata}{Universidad Nacional de La Plata}
\newAff{\atTuebingen}{Eberhard Karls Universit\"at T\"ubingen} 
\newAff{\atFirenze}{Universit\`a degli Studi di Firenze} 
\newAff{\atUCLA}{University of California, Los Angeles}
\newAff{\atWhittier}{Whittier College}
\newAff{\atUNF}{University of North Florida}
\newAff{\atFermilab}{Fermi National Accelerator Laboratory}
\newAff{\atVT}{Virginia Polytechnic Institute and State University} 
\newAff{\atIPHC}{Institut Pluridisciplinaire Hubert Curien (IPHC)}
\newAff{\atSaitama}{Saitama University}
\newAff{\atICTPSAIFR}{International Center for Theoretical Physics -- South American Institute for Fundamental Research} 
\newAff{\atICRAR}{International Centre for Radio Astronomy Research, Curtin University}
\newAff{\atBerkeley}{University of California, Berkeley} 
\onlyastrophysical{\newAff{\atWurzburg}{Julius-Maximilians-Universit\"at W\"urzburg}}
\newAff{\atKonan}{Konan University}
\newAff{\atJHU}{Johns Hopkins University}
\newAff{\atQM}{Queen Mary University of London}
\newAff{\atColumbia}{Columbia University} 
\newAff{\atChiba}{Chiba University}
\newAff{\atLBL}{Lawrence Berkeley National Laboratory}
\newAff{\atHumboldt}{Humboldt-Universit\"at zu Berlin} 
\newAff{\atGutenberg}{Johannes Gutenberg-Universit\"at Mainz}
\newAff{\atCERN}{CERN} 
\onlyfundamental{\newAff{\atIndiana}{Indiana University}}
\newAff{\atAPC}{Laboratoire AstroParticule et Cosmologie} 
\newAff{\atNebraska}{University of Nebraska-Lincoln}
\newAff{\atUMBC}{University of Maryland, Baltimore County}
\newAff{\atDrexel}{Drexel University}
\newAff{\atINFNPisa}{Istituto Nazional di Fisica Nucleare (INFN), Sezione di Pisa}
\newAff{\atUAH}{University of Alabama in Huntsville}
\newAff{\atKU}{University of Kansas}
\newAff{\atHawaii}{University of Hawaii, Manoa}
\newAff{\atMIT}{Massachusetts Institute of Technology} 
\newAff{\atSLAC}{SLAC National Accelerator Lab}
\newAff{\atTrieste}{Universit\`a degli Studi di Trieste}
\newAff{\atASU}{Arizona State University}
\newAff{\atSapienza}{Sapienza – Universit\`a di Roma}
\newAff{\atUWRF}{University of Wisconsin-River Falls}
\newAff{\atINFNPadova}{Istituto Nazional di Fisica Nucleare (INFN), Sezione di Padova}
\onlyfundamental{\newAff{\atWurzburg}{Julius-Maximilians-Universit\"at W\"urzburg}}
\newAff{\atPuebla}{Benem\'erita Universidad Aut\'onoma de Puebla} 
\newAff{\atGranada}{Universidad de Granada}
\onlyastrophysical{\newAff{\atKings}{King's College London}}
\newAff{\atQueens}{Queen's University}
\newAff{\atMercer}{Mercer University}
\onlyfundamental{\newAff{\atCalPoly}{California Polytechnic State University}}
\newAff{\atMilan}{Universit\`a degli Studi di Milano}
\newAff{\atCinestav}{Centro de Investigaci\'on y de Estudios Avanzados del Instituto Polit\'ecnico Nacional (Cinvestav)}
\newAff{\atMPIMunich}{Max-Planck-Institut fur Physik, M\"unchen}
\newAff{\atGeneva}{Universit\'e de Gen\`eve}
\newAff{\atAlberta}{University of Alberta}
\newAff{\atWeizmann}{Weizmann Institute of Science} 
\newAff{\atAuvergne}{Universit\'e Clermont Auvergne}
\newAff{\atMTU}{Michigan Technological University} 
\newAff{\atMontpellier}{Universit\'e de Montpellier}
\newAff{\atESO}{European Southern Observatory} 
\newAff{\atGT}{Georgia Institute of Technology} 
\newAff{\atCrete}{University of Crete} 
\newAff{\atTorVergata}{Universit\`a degli Studi di Roma Tor Vergata}
\newAff{\atNCBJ}{Naradowe Centrum Bada\'n J\k{a}drowych}
\newAff{\atISSRomania}{Institutul de \cb{S}tiin\cb{t}e Spa\cb{t}iale } 
\newAff{\atWashU}{Washington University in St. Louis}
\newAff{\atJoburg}{University of Johannesburg} 
\onlyastrophysical{\newAff{\atLUTH}{Laboratoire Univers et Th\'eories}  }
\newAff{\atIowa}{University of Iowa} 
\newAff{\atTUM}{Technische Universit\"at M\"unchen} 
\newAff{\atUCSD}{University of California, San Diego}  
\newAff{\atVUAms}{Vrije Universiteit Amsterdam} 
\newAff{\atSKKU}{Sungkyunkwan University (SKKU)}
\newAff{\atMichigan}{University of Michigan, Ann Arbor} 
\newAff{\atLeiden}{Universiteit Leiden}
\newAff{\atBarca}{Universitat de Barcelona} 
\newAff{\atBama}{University of Alabama} 
\newAff{\atWuppertal}{Bergische Universit\"at Wuppertal}
\newAff{\atGroningen}{Rijksuniversiteit Groningen} 
\onlyastrophysical{\newAff{\atSaclay}{Universit\'e Paris-Saclay}}
\newAff{\atNapoli}{Universit\`a degli Studi di Napoli Federico II}
\newAff{\atAnnecy}{Universit\'e Grenoble Alpes, Laboratoire d'Annecy-le-Vieux de Physique Th\'eorique (LAPTh)} 
\newAff{\atHamburg}{Universit\"at Hamburg} 
\newAff{\atPrinceton}{Princeton University}
\newAff{\atBologna}{Universit\`a degli Studi di Bologna}
\newAff{\atGenova}{Universit\`a degli Studi di Genova}
\newAff{\atINFNGenova}{Istituto Nazional di Fisica Nucleare (INFN), Sezione di Genova}
\newAff{\atISSCSIC}{Institute of Space Sciences (IEEC-CSIC)}
\newAff{\atTufts}{Tufts University}
\newAff{\atLaAquila}{Universit\`a degli Studi dell'Aquila}
\newAff{\atLeeds}{University of Leeds}
\newAff{\atvandy}{Vanderbilt University}
\newAff{\atTDLI}{Tsung-Dao Lee Institute}
\newAff{\atSunYetsen}{Sun Yet-sen University}
\newAff{\atNovaGorica}{Univerza v Novi Gorici}
\newAff{\atUNLV}{University of Nevada, Las Vegas}
\newAff{\atLosAlamos}{Los Alamos National Laboratory}
\newAff{\atInnsbruck}{Leopold-Franzens-Universit\"at Innsbruck}

\mbox{Kevork N. Abazajian}\atUCI,
\mbox{Sanjib Kumar Agarwalla}\atIOPB,
\mbox{Juan Antonio Aguilar S\'anchez}\atULB,
\mbox{Marco Ajello}\atClemson,
\mbox{Roberto Aloisio}\atINFN~\atGSSI, 
\mbox{Jaime \'Alvarez-Mu\~niz}\atCompostela,
\mbox{Rafael Alves Batista}\atSP,
\mbox{Hongjun An}\atChungbuk,
\mbox{Karen Andeen}\atMarquette, 
\mbox{Shin'ichiro Ando}\atAmsterdam,
\mbox{Gisela Anton}\atErlangen,
\mbox{Ignatios Antoniadis}\atSorbonne~\atBerne,
\mbox{Katsuaki Asano}\atTokyo,
\mbox{Katie Auchettl}\atNBI,
\mbox{Jan Auffenberg}\atRWTHAachen,
\mbox{Hugo Ayala}\atPSU,
\mbox{Xinhua Bai}\atSDSM,
\mbox{Gabriela Barenboim}\atValencia,
\mbox{Vernon Barger}\atUW,
\mbox{Imre Bartos}\atFlorida,
\mbox{Steve W. Barwick}\atUCI,
\mbox{John Beacom}\atOSU,
\mbox{James J. Beatty}\atOSU,
\mbox{Nicole F. Bell}\atMelbourne,
\mbox{Jos\'e Bellido}\atAdelaide,
\mbox{Segev BenZvi}\atRochester,
\mbox{Douglas R. Bergman}\atUtah,
\mbox{Jos\'e Bernab\'eu}\atValencia,
\mbox{Elisa Bernardini}\atPadova~\atDESYZeuthen,
\mbox{Mario Bertaina}\atTurin,
\mbox{Gianfranco Bertone}\atAmsterdam,
\mbox{Peter F. Bertone}\atNASAMarshall,
\mbox{Francesca Bisconti}\atINFNTurin,
\mbox{Jonathan Biteau}\atIPNO,
\mbox{Erik Blaufuss}\atUMD,
\mbox{Summer Blot}\atDESYZeuthen,
\mbox{Julien Bolmont}\atINNPPP,
\mbox{Zeljka Bosnjak}\atZagreb,
\mbox{Olga Botner}\atUppsala,
\mbox{Federica Bradascio}\atDESYZeuthen,
\mbox{Esra Bulbul}\atCFA,
\mbox{Alexander Burgman}\atUppsala,
\mbox{Francesco Cafagna}\atINFNBari,
\mbox{Regina Caputo}\atGoddard,
\mbox{M. Carmen Carmona-Benitez}\atPSU,
\mbox{Rossella Caruso}\atCatania,
\mbox{Marco Casolino}\atINFN,
\mbox{Karem Pe\~nal\'o Castillo}\atFSU,
\mbox{Silvia Celli}\atMPI,
\onlyastrophysical{\mbox{S. Bradley Cenko}\atGoddard,}
\mbox{Andrew Chen}\atWitwatersrand,
\mbox{Yaocheng Chen}\atNTU,
\mbox{Talai Mohamed Cherif}\atAnnaba,
\mbox{Nafis Rezwan Khan Chowdhury}\atValencia,
\mbox{Eugene M. Chudnovsky}\atCUNY,
\mbox{Brian A. Clark}\atOSU,
\mbox{Pablo Correa}\atVUB,
\mbox{Doug F. Cowen}\atPSU,
\mbox{Paschal Coyle}\atMarseille, 
\mbox{Linda Cremonesi}\atUCL,
\mbox{Jane Lixin Dai}\atHongKong, 
\mbox{Basudeb Dasgupta}\atTata,
\mbox{Andr\'e de Gouv\^ea}\atNW,
\mbox{Sijbrand de Jong}\atNijmegen~\atNikhef,
\mbox{Simon De Kockere}\atVUB,
\mbox{Jo\~ao R. T. de Mello Neto}\atRio,
\mbox{Luiz de Viveiros}\atPSU,
\mbox{Krijn D. de Vries}\atVUB,
\onlyastrophysical{\mbox{Gwenha\"el de Wasseige}\atAPC,}
\mbox{Valentin Decoene}\atIAP,
\mbox{Peter B. Denton}\atBNL,
\mbox{Tyce DeYoung}\atMSU,
\mbox{Rebecca Diesing}\atChicago,
\mbox{Markus Dittmer}\atMunster,
\mbox{Caterina Doglioni}\atLund,
\mbox{Klaus Dolag}\atLMU,
\mbox{Michele Doro}\atPadova,
\mbox{Michael A. DuVernois}\atUW,
\mbox{Toshikazu Ebisuzaki}\atRIKEN,
\onlyfundamental{\mbox{John Ellis}\atKings,}
\mbox{Rikard Enberg}\atUppsala, 
\mbox{Ralph Engel}\atKIT,
\mbox{Johannes Eser}\atColoradoMines,
\mbox{Arman Esmaili}\atPontificiaRio,
\mbox{Ke Fang}\atStanford,
\mbox{Jonathan L. Feng}\atUCI,
\mbox{Gustavo Figueiredo}\atOKState,
\mbox{George Filippatos}\atColoradoMines,
\mbox{Chad Finley}\atStockholm,
\mbox{Derek Fox}\atPSU,
\mbox{Anna Franckowiak}\atDESYZeuthen,
\mbox{Elizabeth Friedman}\atUMD,
\mbox{Toshihiro Fujii}\atKyoto,
\mbox{Daniele Gaggero}\atUAM,
\mbox{Alberto M. Gago}\atPeru,
\mbox{Thomas Gaisser}\atDelaware,
\mbox{Shan Gao}\atDESYZeuthen,
\mbox{Carlos Garc\'ia Canal}\atLaPlata,
\mbox{Daniel Garc\'ia-Fern\'andez}\atDESYZeuthen,
\mbox{Simone Garrappa}\atDESYZeuthen,
\mbox{Maria Vittoria Garzelli}\atTuebingen~\atFirenze,
\mbox{Graciela B. Gelmini}\atUCLA,
\mbox{Christian Glaser}\atUCI,
\mbox{Allan Hallgren}\atUppsala,
\mbox{Jordan C. Hanson}\atWhittier,
\mbox{Andreas Haungs}\atKIT,
\mbox{John W. Hewitt}\atUNF,
\mbox{Jannik Hofest\"adt}\atErlangen,
\mbox{Kara Hoffman}\atUMD,
\mbox{Benjamin Hokanson-Fasig}\atUW,
\mbox{Dan Hooper}\atFermilab~\atChicago,
\mbox{Shunsaku Horiuchi}\atVT,
\mbox{Feifei Huang}\atIPHC,
\mbox{Patrick Huber}\atVT, 
\mbox{Tim Huege}\atKIT,
\mbox{Kaeli Hughes}\atChicago,
\mbox{Naoya Inoue}\atSaitama,
\mbox{Susumu Inoue}\atRIKEN,
\mbox{Fabio Iocco}\atICTPSAIFR,
\mbox{Kunihito Ioka}\atKyoto,
\mbox{Clancy W. James}\atICRAR,
\mbox{Eleanor Judd}\atBerkeley,
\mbox{Daniel Kabat}\atCUNY,
\onlyastrophysical{\mbox{Matthias Kadler}\atWurzburg,}
\mbox{Fumiyoshi Kajino}\atKonan,
\mbox{Takaaki Kajita}\atTokyo,
\mbox{Marc Kamionkowski}\atJHU,
\mbox{Alexander Kappes}\atMunster,
\mbox{Dimitra Karabali}\atCUNY,
\mbox{Timo Karg}\atDESYZeuthen,
\mbox{Teppei Katori}\atQM,
\mbox{Uli F. Katz}\atErlangen,
\onlyfundamental{\mbox{Norita Kawanaka}\atKyoto,}
\mbox{Azadeh Keivani}\atColumbia,
\mbox{John L. Kelley}\atUW,
\mbox{Myoungchul Kim}\atChiba,
\mbox{Shigeo S. Kimura}\atPSU,
\mbox{Spencer Klein}\atLBL,
\mbox{Stefan Klepser}\atDESYZeuthen, 
\mbox{David Koke}\atMunster,
\mbox{Hermann Kolanoski}\atHumboldt, 
\mbox{Lutz K\"opke}\atGutenberg,
\mbox{Joachim Kopp}\atGutenberg~\atCERN,
\mbox{Claudio Kopper}\atMSU,
\mbox{Jason Koskinen}\atNBI,
\onlyfundamental{\mbox{V. Alan Kosteleck\'{y}}\atIndiana,}
\mbox{Dmitriy Kostunin}\atDESYZeuthen,
\mbox{Antoine Kouchner}\atAPC,
\mbox{Ilya Kravchenko}\atNebraska,
\mbox{John Krizmanic}\atUMBC,
\mbox{Naoko Kurahashi Neilson}\atDrexel,
\mbox{Michael Kuss}\atINFNPisa,
\mbox{Evgeny Kuznetsov}\atUAH,
\mbox{Ranjan Laha}\atCERN,
\mbox{Uzair Abdul Latif}\atKU,
\mbox{John G. Learned}\atHawaii,
\mbox{Jean-Philippe Lenain}\atSorbonne,
\mbox{Rebecca K. Leane}\atMIT,
\mbox{Shirley Weishi Li}\atSLAC,
\mbox{Lu Lu}\atChiba,
\mbox{Francesco Longo}\atTrieste,
\mbox{Andrew Ludwig}\atChicago,
\mbox{Cecilia Lunardini}\atASU,
\mbox{Paolo Lipari}\atSapienza,
\mbox{James Madsen}\atUWRF,
\mbox{Keiichi Mase}\atChiba,
\mbox{Manuela Mallamaci}\atINFNPadova,
\mbox{Karl Mannheim}\atWurzburg,
\mbox{Danny Marfatia}\atHawaii,
\mbox{Raffaella Margutti}\atNW,
\mbox{Cristian Jes\'us Lozano Mariscal}\atMunster,
\mbox{Szabolcs Marka}\atColumbia,
\mbox{Olivier Martineau-Huynh}\atINNPPP,
\mbox{Oscar Mart\'inez-Bravo}\atPuebla,
\mbox{Manuel Masip}\atGranada,
\mbox{Nikolaos E. Mavromatos}\atKings,
\mbox{Arthur B. McDonald}\atQueens,
\mbox{Frank McNally}\atMercer,
\mbox{Olga Mena}\atValencia,
\mbox{Kevin-Druis Merenda}\atColoradoMines,
\mbox{Philipp Mertsch}\atRWTHAachen,
\mbox{Peter M\'esz\'aros}\atPSU,
\onlyfundamental{\mbox{Matthew Mewes}\atCalPoly,}
\mbox{Hisakazu Minakata}\atTokyo,
\mbox{Nestor Mirabal}\atGoddard,
\mbox{Lino Miramonti}\atMilan,
\mbox{Omar G. Miranda}\atCinestav,
\mbox{Razmik Mirzoyan}\atMPIMunich,
\mbox{John W. Mitchell}\atGoddard,
\mbox{Irina Mocioiu}\atPSU,
\mbox{Teresa Montaruli}\atGeneva,
\mbox{Maria Elena Monzani}\atSLAC,
\mbox{Roger Moore}\atAlberta,
\mbox{Shigehiro Nagataki}\atRIKEN,
\mbox{Masayuki Nakahata}\atTokyo,
\mbox{Jiwoo Nam}\atNTU,
\mbox{Kenny C. Y. Ng}\atWeizmann,
\mbox{Ryan Nichol}\atUCL,
\mbox{Valentin Niess}\atAuvergne,
\mbox{David F. Nitz}\atMTU,
\mbox{Samaya Nissanke}\atAmsterdam,
\mbox{Eric Nuss}\atMontpellier,
\mbox{Eric Oberla}\atChicago,
\mbox{Stefan Ohm}\atDESYZeuthen,
\mbox{Kouji Ohta}\atKyoto,
\mbox{Foteini Oikonomou}\atESO,
\mbox{Roopesh Ojha}\atUMBC~\atGoddard,
\mbox{Nepomuk Otte}\atGT,
\mbox{Timothy A. D. Paglione}\atCUNY,
\mbox{Sandip Pakvasa}\atHawaii,
\mbox{Andrea Palladino}\atDESYZeuthen,
\mbox{Sergio Palomares-Ruiz}\atValencia,
\mbox{Vasiliki Pavlidou}\atCrete,
\mbox{Carlos P\'erez de los Heros}\atUppsala,
\mbox{Christopher Persichilli}\atUCI,
\mbox{Piergiorgio Picozza}\atINFN~\atTorVergata,
\mbox{Zbigniew Plebaniak}\atNCBJ,
\mbox{Vlad Popa}\atISSRomania,
\mbox{Steven Prohira}\atOSU,
\mbox{Bindu Rani}\atGoddard,
\mbox{Brian Flint Rauch}\atWashU,
\mbox{Soebur Razzaque}\atJoburg,
\onlyastrophysical{\mbox{Nicolas Renault-Tinacci}\atLUTH,}
\mbox{Mary Hall Reno}\atIowa,
\mbox{Elisa Resconi}\atTUM,
\mbox{Marco Ricci}\atINFN,
\mbox{Jarred M. Roberts}\atUCSD,
\mbox{Nicholas L. Rodd}\atBerkeley~\atLBL,
\onlyfundamental{\mbox{Werner Rodejohann}\atMPI,}
\mbox{Juan Rojo}\atVUAms,
\mbox{Carsten Rott}\atSKKU,
\mbox{Iftach Sadeh}\atDESYZeuthen,
\mbox{Benjamin R. Safdi}\atMichigan,
\mbox{Naoto Sakaki}\atRIKEN,
\onlyfundamental{\mbox{David Saltzberg}\atUCLA,}
\mbox{Jordi Salvad\'o}\atBarca,
\mbox{Dorothea Samtleben}\atLeiden,
\mbox{Marcos Santander}\atBama,
\mbox{Fred Sarazin}\atColoradoMines,
\mbox{Konstancja Satalecka}\atDESYZeuthen,
\mbox{Michael Schimp}\atWuppertal,
\mbox{Olaf Scholten}\atGroningen,
\mbox{Harm Schoorlemmer}\atMPI,
\onlyastrophysical{\mbox{Frank G. Schr\"oder}\atDelaware,}
\onlyastrophysical{\mbox{Fabian Sch\"ussler}\atSaclay,}
\mbox{Sergio J. Sciutto}\atLaPlata,
\mbox{Valentina Scotti}\atNapoli,
\mbox{David Seckel}\atDelaware,
\mbox{Pasquale D. Serpico}\atAnnecy,
\mbox{Shashank Shalgar}\atNBI,
\mbox{Jerry Shiao}\atNTU,
\onlyastrophysical{\mbox{Ankur Sharma}\atINFNPisa,}
\mbox{Kenji Shinozaki}\atTurin,
\mbox{Ian M. Shoemaker}\atVT,
\mbox{G\"unter Sigl}\atHamburg,
\mbox{Lorenzo Sironi}\atColumbia,
\mbox{Tracy R. Slatyer}\atMIT,
\mbox{Radomir Smida}\atChicago,
\mbox{Alexei Yu Smirnov}\atMPI,
\mbox{Jorge F. Soriano}\atCUNY,
\mbox{Daniel Southall}\atChicago,
\mbox{Glenn Spiczak}\atUWRF,
\mbox{Anatoly Spitkovsky}\atPrinceton,
\mbox{Maurizio Spurio}\atBologna,
\mbox{Juliana Stachurska}\atDESYZeuthen, 
\mbox{Krzysztof Z. Stanek}\atOSU,
\mbox{Floyd Stecker}\atGoddard,
\mbox{Christian Stegmann}\atDESYZeuthen,
\mbox{Robert Stein}\atDESYZeuthen,
\mbox{Anna M. Suliga}\atNBI,
\mbox{Greg Sullivan}\atUMD,
\mbox{Jacek Szabelski}\atNCBJ,
\onlyfundamental{\mbox{Ignacio Taboada}\atGT,} 
\mbox{Yoshiyuki Takizawa}\atRIKEN,
\mbox{Mauro Taiuti}\atGenova~\atINFNGenova,
\mbox{Irene Tamborra}\atNBI,
\mbox{Xerxes Tata}\atHawaii,
\mbox{Todd A. Thompson}\atOSU,
\mbox{Charles Timmermans}\atNijmegen~\atNikhef,
\mbox{Kirsten Tollefson}\atMSU,
\mbox{Diego F. Torres}\atISSCSIC,
\mbox{Jorge Torres}\atOSU,
\mbox{Simona Toscano}\atULB,
\mbox{Delia Tosi}\atUW,
\mbox{Mat\'ias Tueros}\atLaPlata,
\mbox{Sara Turriziani}\atRIKEN,
\mbox{Elisabeth Unger}\atUppsala,
\mbox{Michael Unger}\atKIT,
\mbox{Martin Unland Elorrieta}\atMunster,
\mbox{Jos\'e Wagner Furtado Valle}\atValencia,
\mbox{Lawrence Wiencke}\atColoradoMines,
\mbox{Nick van Eijndhoven}\atVUB,
\mbox{Jakob van Santen}\atDESYZeuthen,
\mbox{Arjen van Vliet}\atDESYZeuthen,
\mbox{Justin Vandenbroucke}\atUW, 
\mbox{Gary S. Varner}\atHawaii,
\mbox{Tonia Venters}\atGoddard,
\mbox{Matthias Vereecken}\atVUB,
\mbox{Alex Vilenkin}\atTufts,
\mbox{Francesco L. Villante}\atLaAquila, 
\mbox{Aaron Vincent}\atQueens,
\mbox{Martin Vollmann}\atTUM,
\mbox{Philip von Doetinchem}\atHawaii,
\mbox{Alan A. Watson}\atLeeds,
\mbox{Eli Waxman}\atWeizmann,
\mbox{Thomas Weiler}\atvandy,
\mbox{Christoph Welling}\atDESYZeuthen,
\mbox{Nathan Whitehorn}\atUCLA,
\mbox{Dawn R. Williams}\atBama,
\mbox{Walter Winter}\atDESYZeuthen,
\mbox{Hubing Xiao}\atINFNPadova,
\mbox{Donglian Xu}\atTDLI,
\mbox{Tokonatsu Yamamoto}\atKonan,
\mbox{Lili Yang}\atSunYetsen,
\mbox{Gaurang Yodh}\atUCI,
\mbox{Shigeru Yoshida}\atChiba,
\mbox{Tianlu Yuan}\atUW,
\mbox{Danilo Zavrtanik}\atNovaGorica,
\mbox{Arnulfo Zepeda}\atCinestav,
\mbox{Bing Zhang}\atUNLV,
\mbox{Hao Zhou}\atLosAlamos,
\mbox{Anne Zilles}\atIAP,
\mbox{Stephan Zimmer}\atInnsbruck,
\mbox{Juan de Dios Zornoza}\atValencia,
\mbox{Renata Zukanovich Funchal}\atSP,
\mbox{~and~Juan Z\'u\~niga}\atValencia
\linebreak

\affiliations